\documentclass[fleqn,usenatbib]{mnras}

\usepackage{newtxtext,newtxmath}

\usepackage[T1]{fontenc}

\DeclareRobustCommand{\VAN}[3]{#2}
\let\VANthebibliography\thebibliography
\def\thebibliography{\DeclareRobustCommand{\VAN}[3]{##3}\VANthebibliography}

\usepackage{graphicx}	
\usepackage{amsmath}	

\usepackage[normalem]{ulem}
\useunder{\uline}{\ul}{}

\title[Cold front in RXJ2014]{A Detailed View of the Large-Scale Sloshing Cold Front in RXJ2014.8-2430}

\author[M. J. Sundquist et al.]{
M. J. Sundquist,
S. A. Walker\thanks{Email: stephen.walker@uah.edu},
M. S. Mirakhor
\\
Department of Physics and Astronomy, The University of Alabama in Huntsville, 301 Sparkman Drive NW, Huntsville, AL 35899, USA
}

\date{}

\pubyear{2015}

\begin{document}
\label{firstpage}
\pagerange{\pageref{firstpage}--\pageref{lastpage}}
\maketitle

\begin{abstract}

We analyze our new 144 ks deep Chandra observation of the sloshing cold front cluster RXJ2014.8-2430. Previous observations of RXJ2014.8-2430 with XMM-Newton shows evidence of a large scale, sloshing cold front around 800 kpc away from the cluster core. Previous shallow Chandra data also shows evidence of two younger cold fronts closer to the core. Our new deeper Chandra data allow us to analyze the fine, small scale structure of these three cold fronts. Using both beta model subtraction and Gaussian Gradient Magnitude filtering, we confirm the locations of the three cold fronts, as well as discover a large concave structure southeast of the cluster core near the outermost cold front, which could be a large Kelvin-Helmholtz instability or a gas cavity from AGN activity. Analyzing the three cold fronts, we measure the widths of the cold fronts and find them to be consistent with or lower than the Coulomb mean free paths within error, signifying that diffusion is suppressed across the cold fronts. If the concave feature is the inner rim of a cavity, we find that it has a radius in the range 200-330kpc, and would have $PV$ values in the range of $5.7 \times 10^{60}$  - $2.7 \times 10^{61}$ erg. These values would make it consistent with the some of the most powerful bubbles observed.

\end{abstract}




\section{Introduction}

Galaxy clusters, the largest virialized structures in the universe, form by merging with smaller structures such as galaxies or galaxy groups \citep{Ascasibar2006,Markevitch2007}. The gravitational disturbances from off-axis minor mergers involving a cool core cluster causes the intracluster medium (ICM) to slosh and form cold fronts \citep{Tittley2005}. In simulations cold fronts are simulated to propagate outwards into the outer regions of the cluster, moving in a characteristic spiral pattern caused by the tangential motion of the merging object. At the interface of a cold front, the x-ray brightness and gas density drop sharply, while the temperature rises sharply.

While we observe that these cold fronts rise from the center of the cluster, very few cold fronts have been observed in the regions far outside the cooling radius of the cluster, which is the radius where the cooling time is less than the age of the universe. The difficulty in identifying cold fronts in these outer regions is mainly due to the low surface brightness of the ICM, as well as the small field of view of observations, making it necessary to create mosaics of observations to create a full image of a cluster. Despite the difficulty, a growing sample of cold fronts have been found at around half the virial radius, including the Perseus Cluster \citep{Simionescu2012, Walker2018NatAs},  Abell 2142 \citep{Rossetti2013}, Abell 1763 \citep{Douglass2018} RXJ2014.8-2430 \citep{Walker2014}, Abell 3558 \citep{Mirakhor2023}, and Abell 2029 \citep{Watson2025}.

\begin{figure*}
    \centering
    \hspace{0.45cm}
     \hbox{
    \includegraphics[scale=0.3]{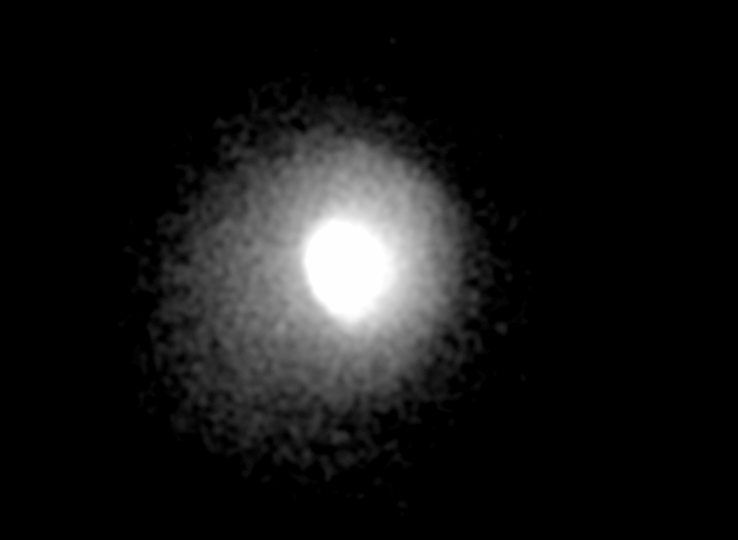}
    \includegraphics[scale=0.3]{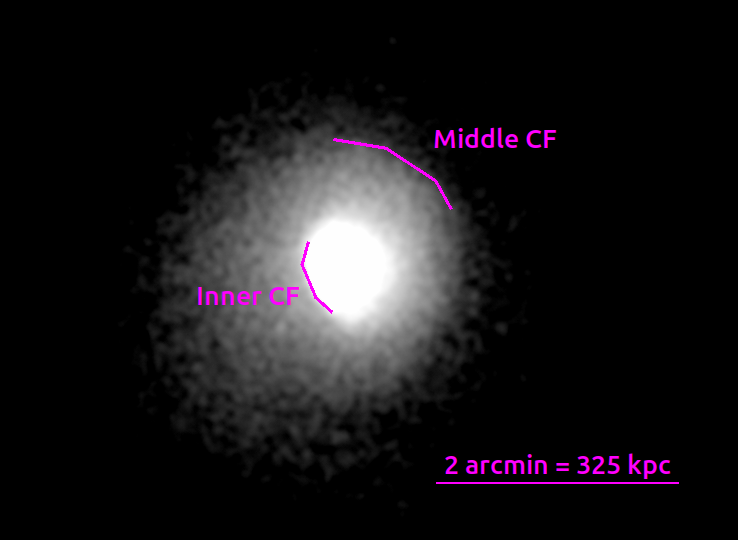}}
    
    \hspace{0.5cm}
    \hbox{
    \includegraphics[scale=0.3845]{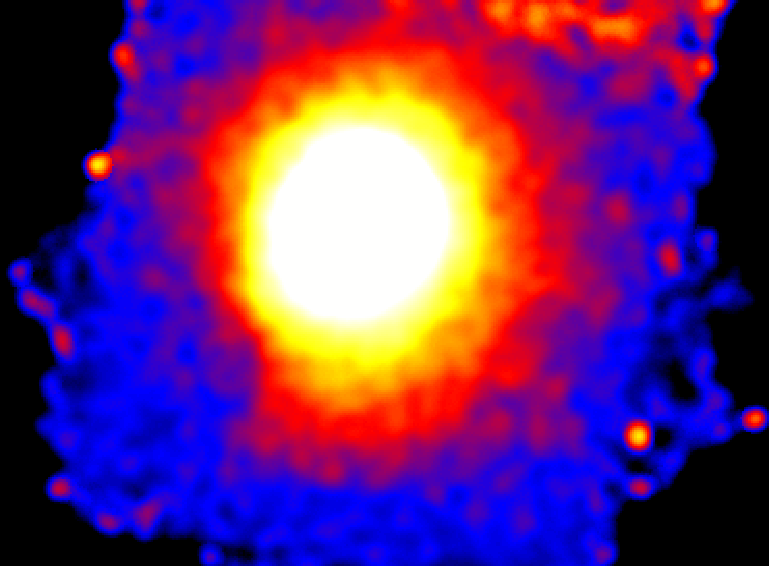}
    \includegraphics[scale=0.3845]{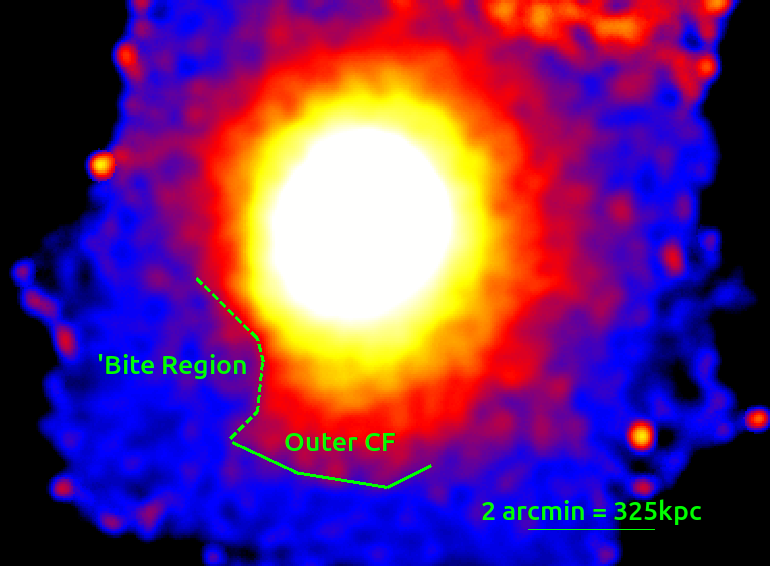}}
    \caption{Exposure corrected, {background subtracted}, point source subtracted images of RXJ2014 in the 0.7-7.0 keV band. \textbf{Top:} Scaled exposure corrected images of the cluster to show the inner two cold fronts. The left image is unlabeled and the right image shows the locations of the inner two cold fronts. \textbf{Bottom:} Scaled exposure corrected images of the cluster to show the outer cold front and the bite region. The left image is unlabeled and the right image shows the locations of both the outermost cold front and the shape of the bite region.}
    \label{diff1}
\end{figure*}

Due to the {shear velocity} of the gases on either side of the cold front interface, there is a potential for Kelvin-Helmholtz instabilities (KHI) to form along the edge. While the cold front generally follows an outward {swirling} pattern, KHI can form as concave regions along the edge. These instabilities can be suppressed, such as by increased magnetic field strengths in the ICM due to magnetic draping \citep{Zuhone2018}, and as such, instabilities are not always present in cold fronts. Large scale cold fronts however are older which gives more time for instabilities to form along the interface. They would also potentially be larger, which would make them easier to detect. There are a growing number of clusters in which we find bite shaped features located only on a single side of the cluster, where it is not clear if they are due to AGN feedback. Examples of these are found in the Perseus Cluster and the Centaurus Cluster \citep{Walker2017}, Abell 1795 \citep{Walker2014a1795}, the Ophiuchus Cluster \citep{Giacintucci2020}, and Abell 2029 \citep{Watson2025}.

A large scale cold front 800 kpc from the cluster core was found in RXJ2014.8-2430 (hereafter RXJ2014), the strongest cool core cluster in the REXCESS galaxy cluster catalog \citep{Walker2014}. This cold front was observed with XMM Newton data and due to the poor spatial resolution of XMM Newton, analysis on the fine structure of this cold front and its width could not be measured. Here, we present our results from our new deep 144ks observation of RXJ2014 [PI: S. A. Walker] with Chandra, whose exceptionally good point spread function allows us to analyze the small scale structure of the large scale cold front. 

For RXJ2014, we use a mass value of $M_{500}$ = $5.4\times10^{14} M_\odot$ \citep{Pratt2010}, and a redshift value of {z = 0.154.} We use a standard $\Lambda$CDM cosmology with $H_o = 70$ km s$^{-1}$ Mpc$^{-1}$, $\Omega$$_{M} = 0.3$ and $\Omega$$_{\Lambda} = 0.7$. All errors are at the 1$\sigma$ level unless otherwise stated.

\begin{figure*}
    \centering
    \includegraphics[scale=0.5]{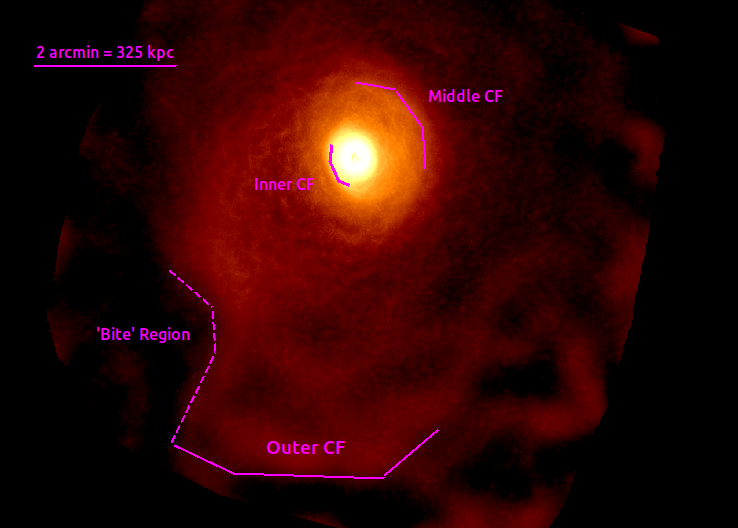}
    \caption{Composite GGM image using three separate GGM filters with different smoothing scales added together. The smoothing scales used were 2, 8, and 32 pixels. We can see the three individual cold fronts in the ICM as well as the concave region, or `bite' region in the southeastern part of the outer ICM.}
    \label{RXJggm}
\end{figure*}

\begin{figure*}
    \centering
    \vbox{
    \hbox{
    \includegraphics[scale=0.35]{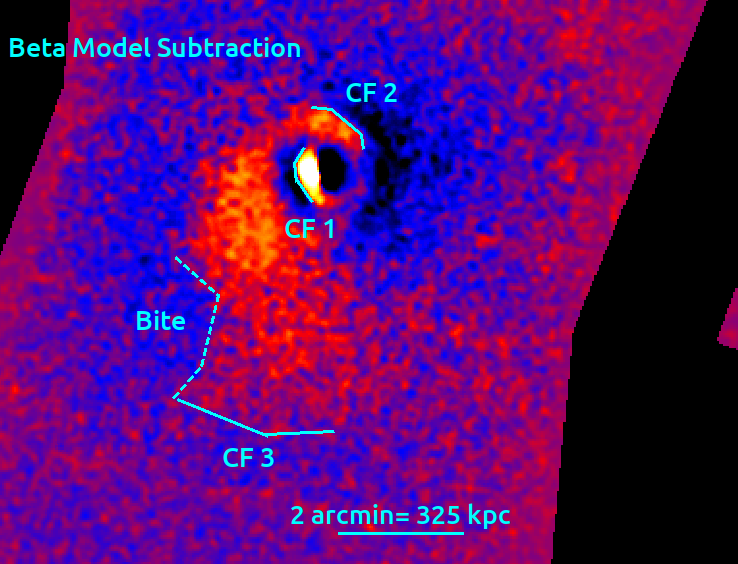}
    \includegraphics[scale=0.31]{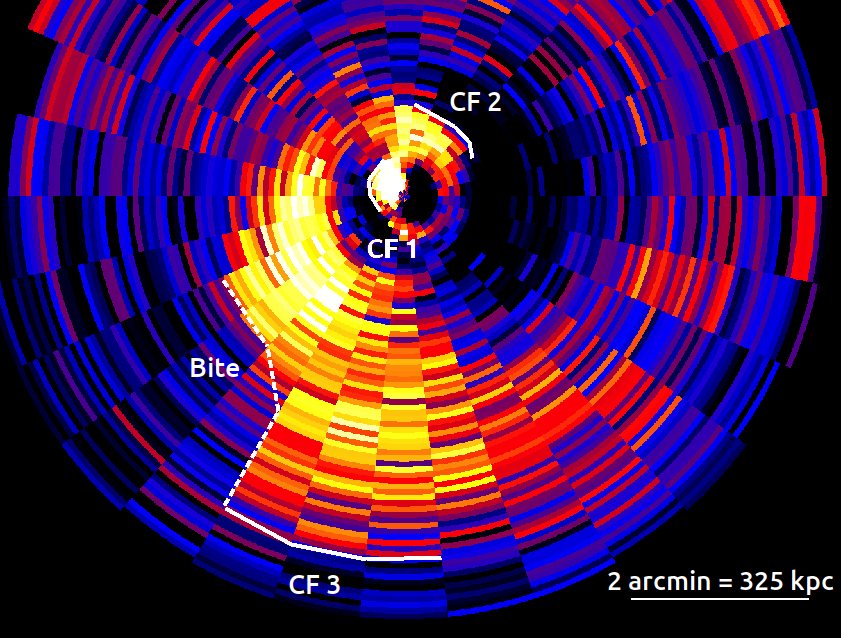}}
    \hbox{
    \includegraphics[scale=0.308]{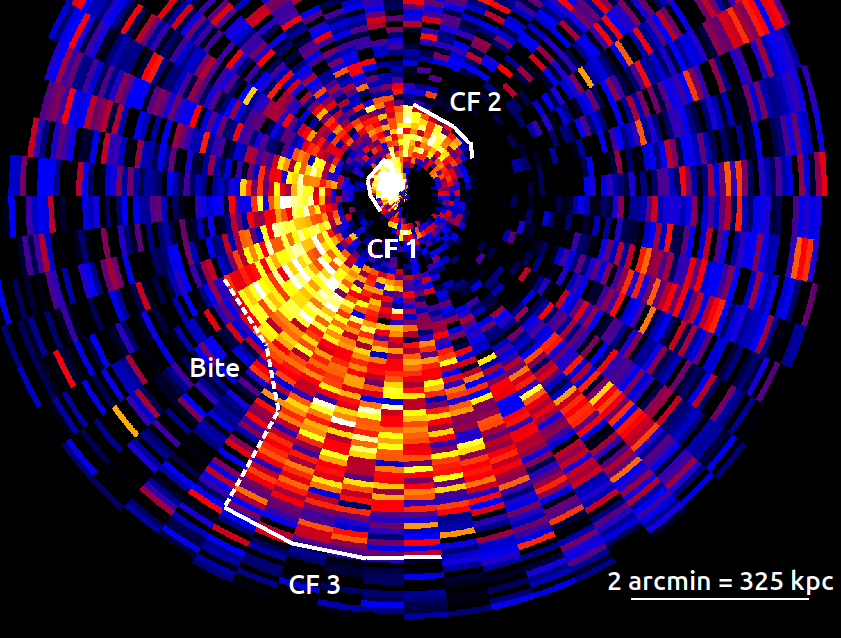}
    \includegraphics[scale=0.308]{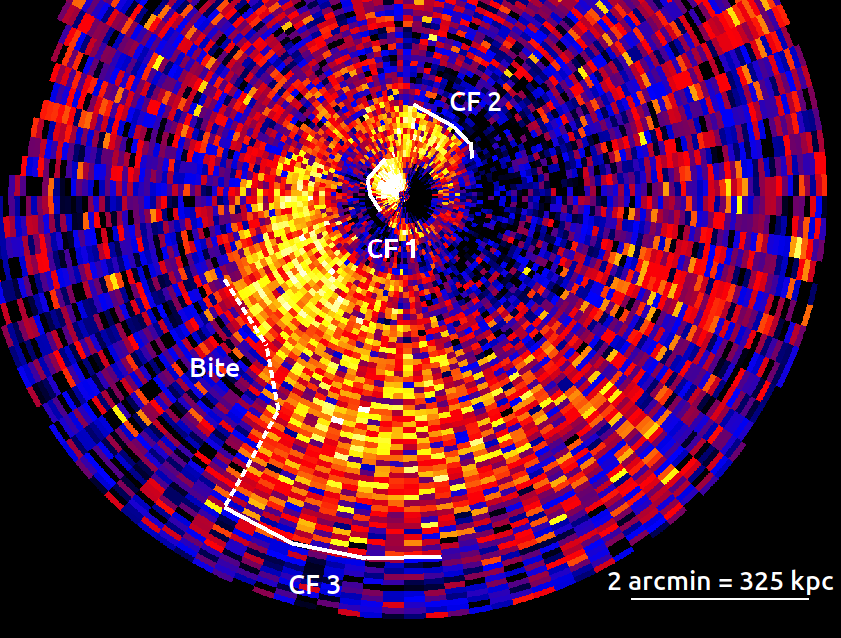}
    }}
    \caption{We use beta model subtraction plots to highlight the structures in the ICM of RXJ2014. Using beta model subtraction with a best fit value of $\beta$ of 0.7, we find that all three cold fronts and the bite sector can be resolved and seen more clearly. Then, we bin regions together into sectors and use the same method. \textbf{Top Left} shows the result of beta model subtraction \textbf{Top Right} uses 30 sectors, \textbf{Bottom Left} uses 60 sectors, and \textbf{Bottom Right} uses 120 sectors. We can identify the shape of the bite in all three images.}
    \label{scatter30_60}
\end{figure*}

\section{Data reduction}
\label{sec: data}

The observations taken of RXJ2014 were performed with Chandra's ACIS-S detector, with a total raw exposure time of around 170 ks. The full list of observations used, the location in the sky of the observation, the date of observation, the exposure time of the observation, and the cleaned exposure times are listed in Table \ref{obs}.

\begin{table*}
\centering
\begin{tabular}{llllll}
\hline
\multicolumn{1}{|l}{\textbf{Obs ID}} & \textbf{RA} & \textbf{Dec} & \textbf{Date} & \textbf{Exp Time (ks)} & \multicolumn{1}{l|}{\textbf{Cleaned Time (ks)}} \\ \hline
11757 & 20 14 51.70 & -24 30 22.90 & 2009 Aug 25  & 19.91 & 18.01 \\
25466 & 20 14 52.01 & -24 31 56.18 & 2023 Jul 16  & 15.90 & 10.74  \\
25838 & 20 14 52.01 & -24 31 56.18 & 2022 Aug 04  & 28.73 & 24.06 \\
25839 & 20 14 52.01 & -24 31 56.18 & 2022 Aug 08  & 17.81 & 11.83 \\
25840 & 20 14 52.01 & -24 31 56.18 & 2021 Dec 03  & 16.11 & 13.21 \\
25841 & 20 14 52.01 & -24 31 56.18 & 2022 Aug 27 & 29.72 & 22.34 \\
26221 & 20 14 52.01 & -24 31 56.18 & 2021 Dec 04  & 15.90 & 11.95 \\
27253 & 20 14 52.01 & -24 31 56.18 & 2022 Aug 09  & 13.44 & 9.73  \\
27254 & 20 14 52.01 & -24 31 56.18 & 2022 Aug 10  & 13.93 & 10.22 \\
27255 & 20 14 52.01 & -24 31 56.18 & 2023 Jul 31  & 17.06 & 12.70 \\
\hline
\end{tabular}
\caption{List of all observations used in this analysis, with their observation ID's, location of the sky the data was taken, the date of the observation, and both the original exposure time along with the cleaned exposure time.}
\label{obs}
\end{table*}

The data reduction used in the observation analysis was done using the Chandra Interactive Analysis of Observations (CIAO) \citep{Fruscione2006}. The primary data is comprised of photon counts that hit the Chandra ACIS chips during the observation, which {record} the location of photon, energy of the photon, and the time the photon arrived. This original data was then reprocessed using the \textit{chandra\_repro} script. The reprocessed files were then put through a \textit{deflare} routine, which uses \textit{lc\_clean} to identify flares in the observation by finding X-ray counts 3$\sigma$ above or below the average count rate for an observation. Once identified, those times are removed from the data, which left 144 ks of exposure time after cleaning. {To subtract the particle background, we obtain stowed background files.  We then normalize these stowed background files by the count rate in the 9-12 keV band so that they match the actual observation. We then subtract those stowed background images from the images of our cluster.}

Once the observations were cleaned, the individual observations were merged together using \textit{merge\_obs} which both aligns the images up with one another while also creating exposure maps. This was performed for the 0.7-7.0 keV band. The script then creates an exposure corrected image by taking the observation and dividing it by the exposure map.

The observations then needed to have point sources identified, which was done using the \textit{wavdetect} script. The source pixels are then removed and replaced with an average count rate from the surrounding location using \textit{dmfilth}. The exposure corrected image with point sources removed can be seen in Fig \ref{diff1}.


\begin{figure*}
    \centering
    \hbox{
    \includegraphics[scale=0.35]{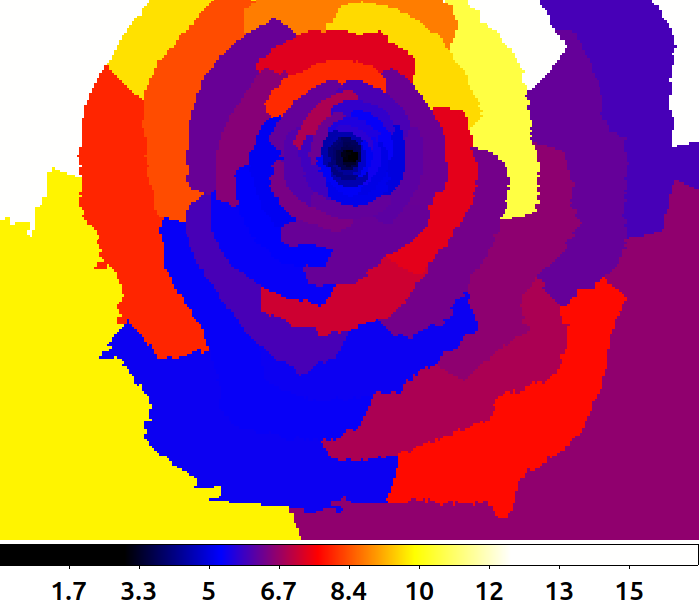}
    \includegraphics[scale=0.35]{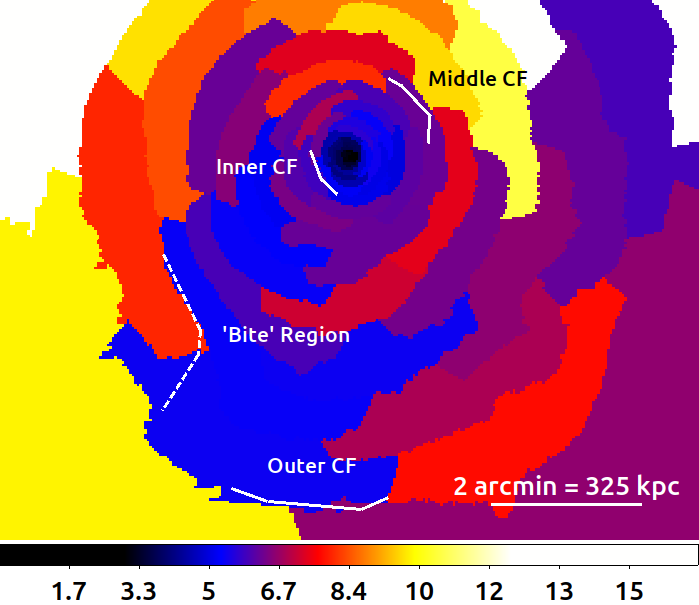}}
    \hbox{
    \includegraphics[scale=0.34]{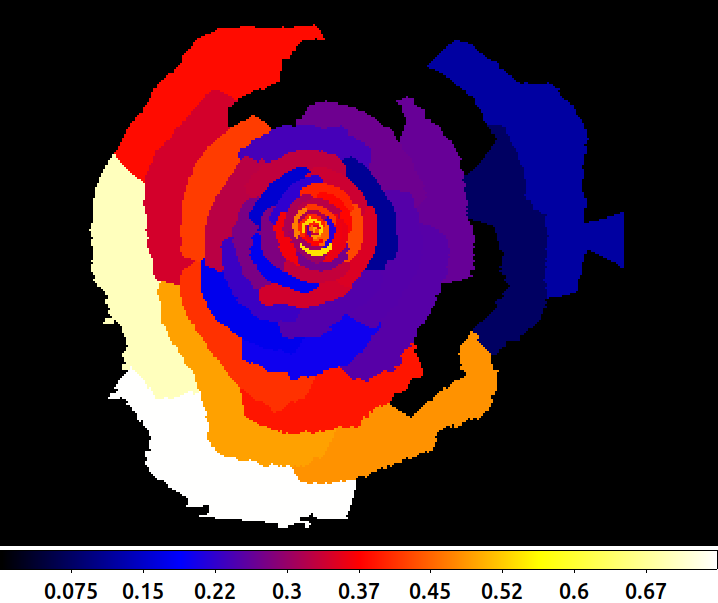}
    \includegraphics[scale=0.34]{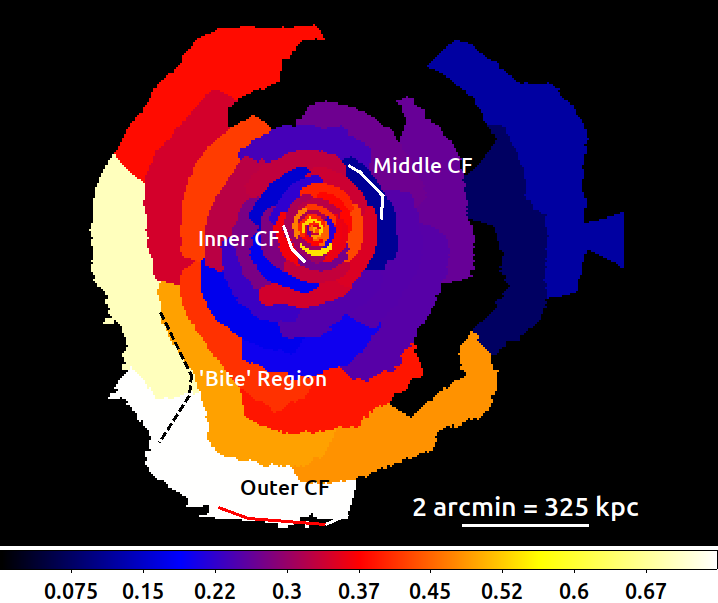}
    }
    \caption{A comparison of the temperature map (Top) and abundance map (Bottom) for a signal to noise of 70. The left hand panels show the maps without labels, and the right hand panels show the plots with the main features in the X-ray image labelled. In both maps, we can identify the swirled structured of the ICM, following the path of cold temperatures and high abundances as they rise from the cluster core, showing the path of the sloshing in the cluster.}
    \label{temp_abun}
\end{figure*}

\begin{figure*}
    \centering
    \hbox{
    \includegraphics[scale=0.55]{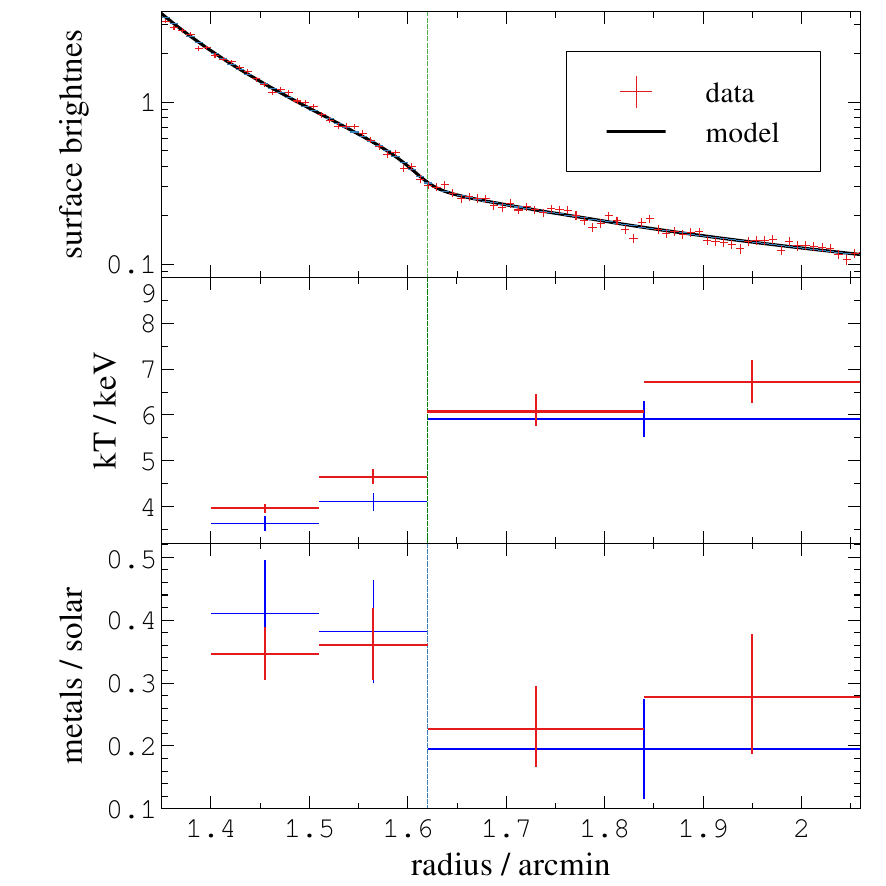}
    \includegraphics[scale=0.34]{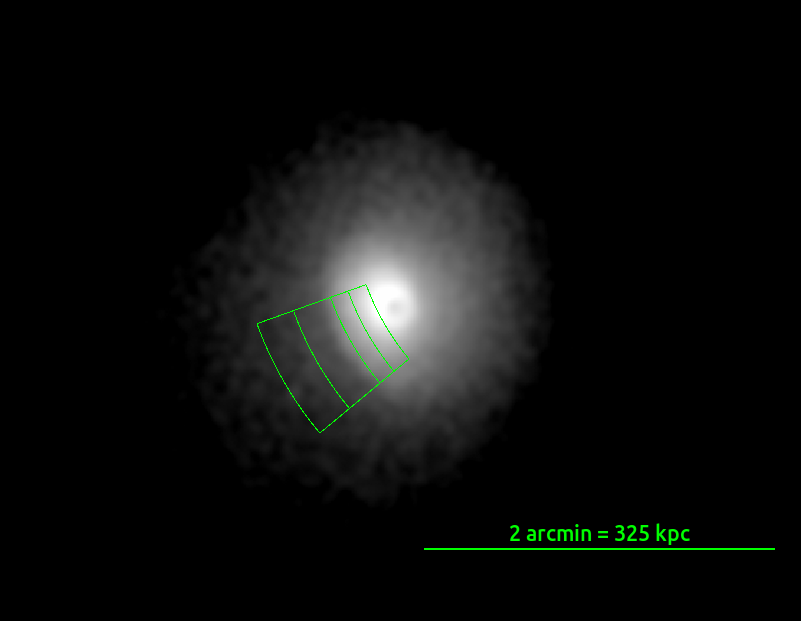}}
    \caption{\textbf{Left:} Surface brightness profile, temperature profile, and abundance profile for the innermost cold front. The green line denotes the radius of the cold front interface. The red data points in the temperature and abundance profiles are projected values, while the blue data points are deprojected. \textbf{Right:} The sector used in the analysis of the innermost cold front.}
    \label{200-220}
\end{figure*}

\section{Image Analysis: Highlighting Edges and Structures}
\label{edges}

The locations of the three cold fronts are shown on the exposure corrected images in Fig \ref{diff1}. We also use the method of Gaussian Gradient Magnitude (GGM) filtering to highlight the gradients in the surface brightness in the data \citep{Sanders2016b, Walker2016}. Different filtering parameters can be used to highlight different scales of changes, with lower parameters highlighting more subtle changes and larger parameters highlighting more broad changes. We run the GGM filter on RXJ2014 using smoothing scales of 2, 4, 8, 16, and 32 pixels. We then create a composite image of the GGM images by adding the GGM images together after applying a scale factor to them. The scale factor makes it so that the more broadly filtered images do not dominate in the center of the cluster. Using the images from the 2, 8, and 32 filter, we create the composite image seen in Fig \ref{RXJggm}. In this image, we can see the three individual cold fronts clearly as well as a large concave structure in the southeastern portion of the image, which we refer to as the `bite' region.

We also use the method of beta model subtraction to highlight the cold fronts and this `bite' region. We create a best-fit model for the cluster and assume that the surface brightness drops with radius as in a beta model. We then subtract the model from our data which will highlight the points that have vastly higher or lower values in the count rate compared to what we expect for a spherically symmetric cluster. We then use the same method, however we bin the cluster into angular regions. {The reason we use the method of binning these regions into larger sectors is to show that cold front and the bite region are robust structures that do not depend on how the data are binned together. Our three sector images have a signal to noise of 18, 13, and 9 per region for the 30 sector, 60 sector, and 120 sector binnings respectively.} These images can be seen in Fig \ref{scatter30_60}. What we notice is that in all the images, this `bite' region can be seen as well as the swirl shape created from sloshing.

\section{Spectral Analysis}
\label{spectral}

We use the method of contour binning to first map out the temperatures in the cluster \citep{Sanders2006contour}. We use a signal to noise ratio of 70 which corresponds to 4900 background subtracted photons per region, creating 74 separate regions. This allows for accurate temperature and metal abundance measurements in each region. For each region, we run \textit{specextract} on each events file, which extracts the spectra from the region as well as the RMF and ARF from each region and each events file. 

To subtract the particle background, we use the same techniques used in \cite{Walker2018NatAs}. The particle background was subtracted from the datasets using stowed backgrounds scaled to match the count rate at energy ranges of 9-12 keV. Then, to model the soft X-ray background, we used the ROSAT All Sky Survey background fields with a circular annulus with inner radius 0.25 degrees and an outer radius of 0.5 degrees away from the cluster center to find the best fit values for the unabsorbed thermal component of the local hot bubble and the absorbed thermal component of the galactic halo. We find the temperature of value of the local hot bubble to be 0.23 keV and the temperature value of the galactic halo to be 0.92 keV. We also use a powerlaw of index of 1.4 to model the cosmic X-ray background, and its normalization value was calculated using \cite{Moretti2003}, using the threshold flux for point source subtraction in our Chandra data of 10$^{-15}$ erg s$^{-1}$.

 The intracluster medium emission was fit using a \textsc{phabs(apec)} model, using a hydrogen column density of $n_{H} = 13.1\times10^{20}$ cm$^{-2}$ \citep{Croston2008}, {an energy range of 0.7-7.0 keV which is the effective energy range of Chandra} and a redshift of {z = 0.154}, and using C statistics in \textsc{xspec} \citep{Arnaud1996}. For each region, we get a temperature value and a metallicity, and the temperature and metallicity maps can be seen in Fig \ref{temp_abun} {with errors in the temperature measurements around 10\% and metallicity measurements around 15\%} . We find that the temperature and metal distributions are highly asymmetric on large scales, with a large cold and metal rich region inside the outer cold front. As well, we can see the characteristic swirl structure in the abundance map, showing the metals following the path of the swirl as it moves outward from the core, which is what we expect to see as a result of sloshing \citep{Roediger2011}. 

\section{Cold Fronts}

Across a cold front, we expect to see a sharp discontinuity in the surface brightness. The model for this discontinuity is given by a broken power law model for the gas density:

\begin{equation}
    n(r) = \begin{cases}
    n_{0}(\frac{r}{r_{cf}})^{\alpha_{1}} & \text{if} \; \; r \leq r_{cf} \\
    \frac{1}{C}n_{0}(\frac{r}{r_{cf}})^{\alpha_{2}} & \text{if} \; \; r > r_{cf}
    \end{cases}
    \label{broken_power_law}
\end{equation}

In equation \ref{broken_power_law}, $\alpha_{1}$ and $\alpha_{2}$ are the power law indices, $n_{0}$ is the density normalization, $r_{cf}$ is the distance to the interface of the cold front, and $C = n_{2}/n_{1}$, where $n_{2}$ is the density beyond the cold front, and $n_{1}$ is the density of the cold front \citep{Mirakhor2023}. This factor of $1/C$ creates a jump, or a sharp decrease in the density profile. As such, we refer to this parameter as the jump parameter.

\begin{figure*}
    \centering
    \hbox{
    \includegraphics[scale=0.55]{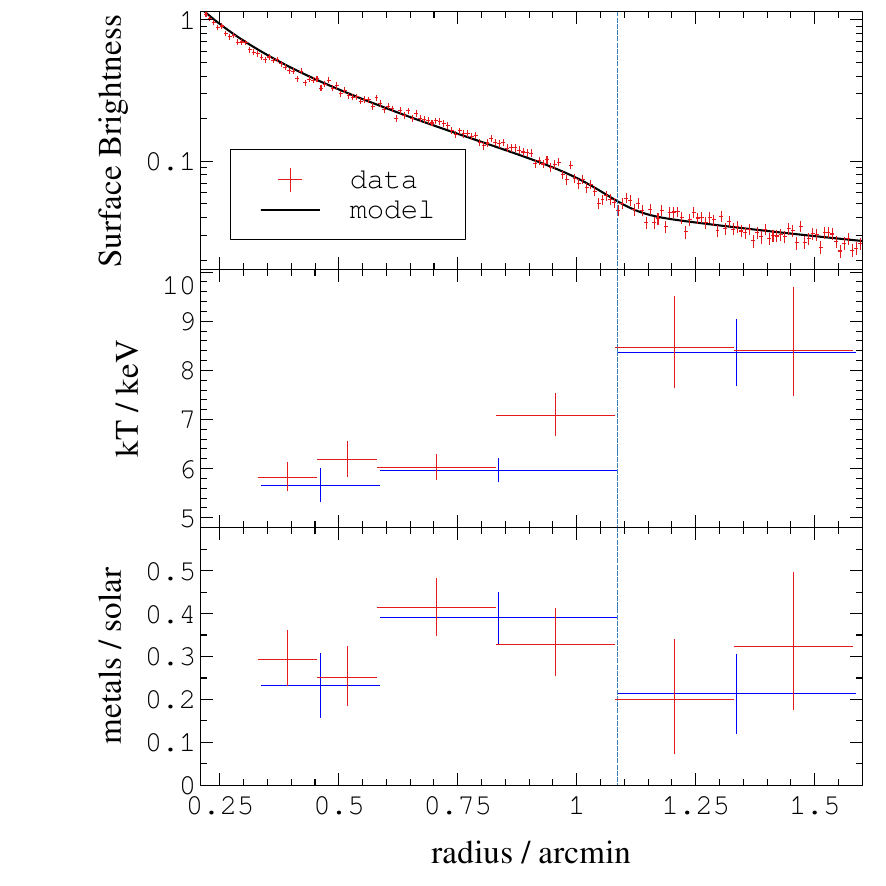}
    \includegraphics[scale=0.365]{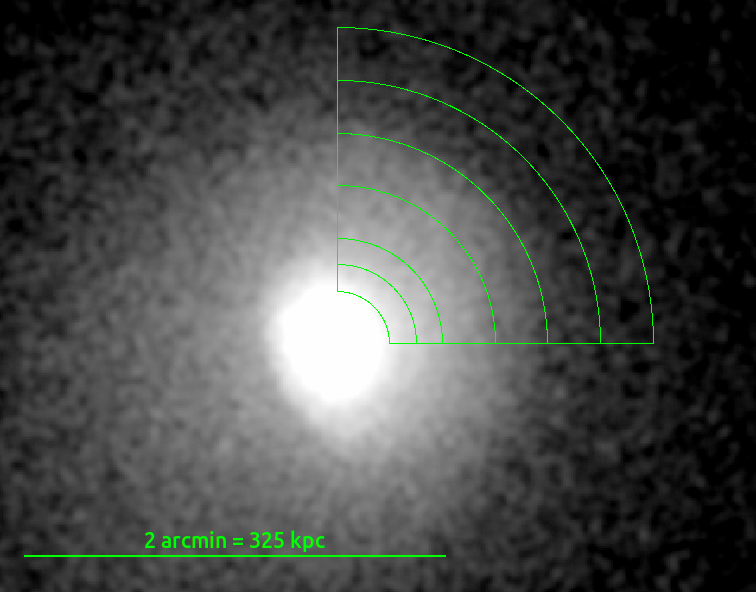}}
    \caption{\textbf{Left:} Surface brightness profile, temperature profile, and abundance profile for the sector angle 0-90 which we determine as the middle cold front. The vertical line denotes the radius of the cold front interface. The red data points in the temperature and abundance profiles are projected values, while the blue data points are deprojected. \textbf{Right:} Sector used in this analysis of the middle cold front.}
    \label{mkreg}
\end{figure*}

\begin{figure*}
    \centering
    \hbox{
    \includegraphics[scale=0.55]{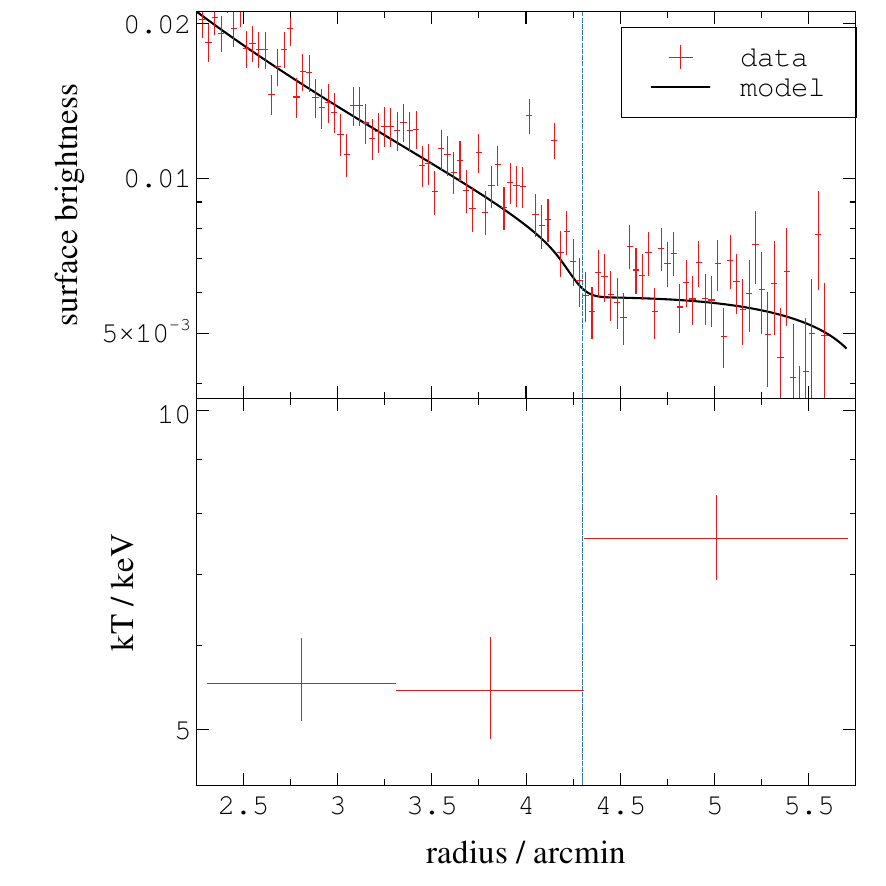}
    \includegraphics[scale=0.38]{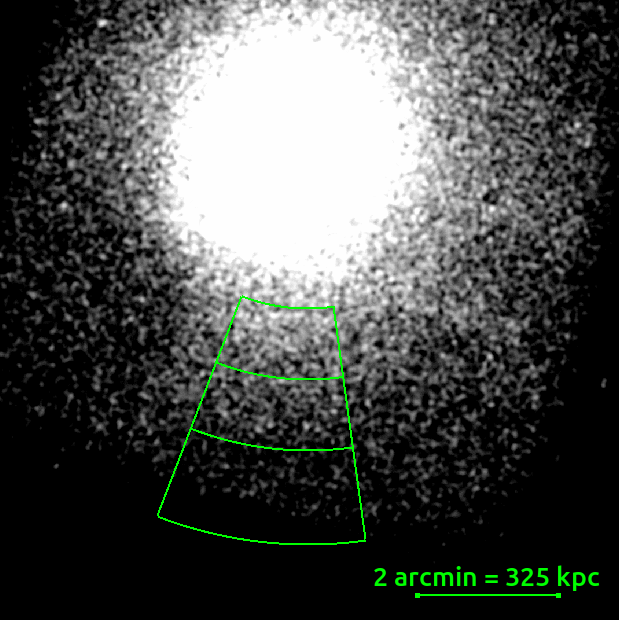}
    }
    \caption{\textbf{Top Left:} Surface brightness profile for the sector angle 249-278 degrees, which covers the outermost cold front. \textbf{Bottom left:} Temperature profile from the same sector. \textbf{Right:} Sector used in this analysis of the outer cold front.}
    \label{250_299}
\end{figure*}

\begin{figure*}
    \centering
    \hbox{
    \includegraphics[scale=0.5]{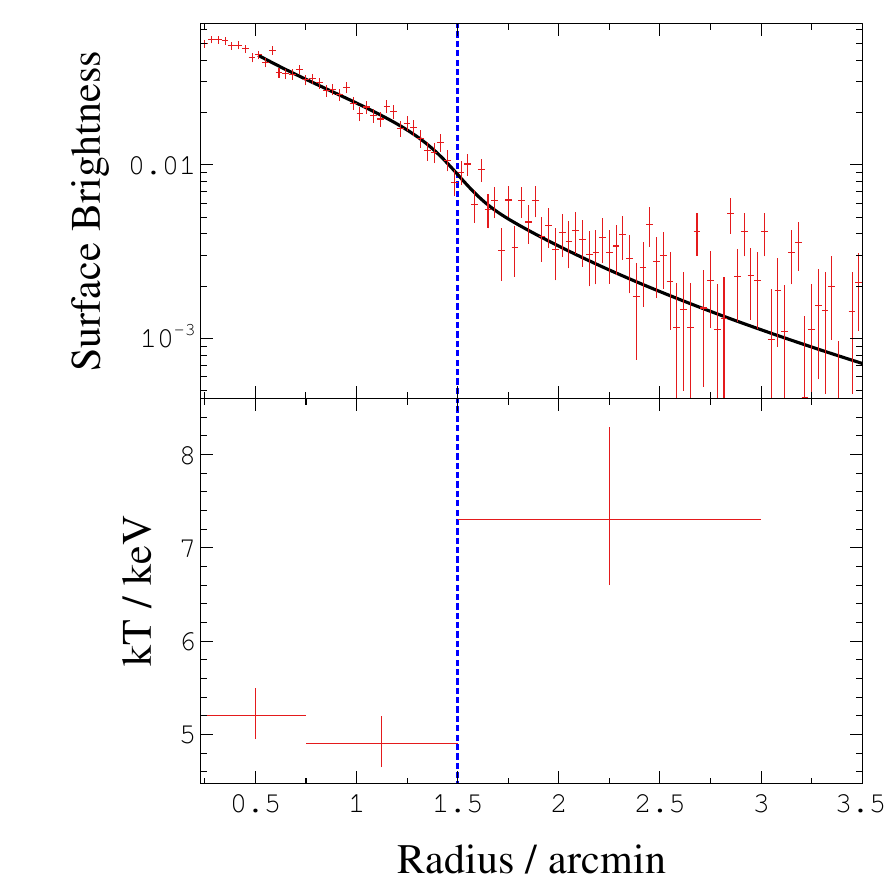}
    \includegraphics[scale=0.398]{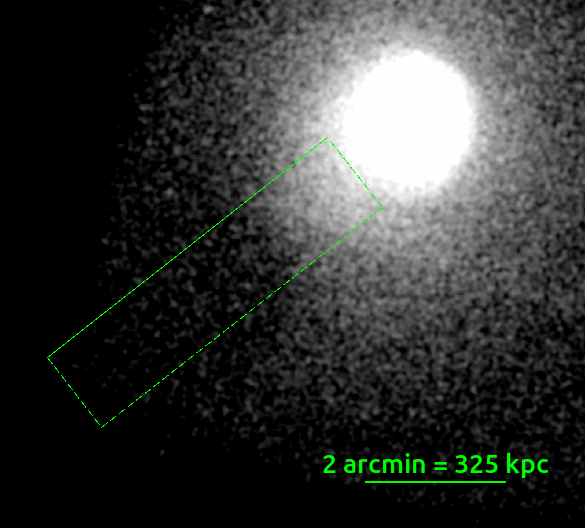}}
    \caption{Left: Surface brightness profile going along the bite, following the region shown in the right hand panel. We fit a broken power law model to the surface brightness profile to find both an estimated value for the jump parameter and the width. 
    We find the jump parameter to be 1.44 for this sector and a width of $17.06_{-15}^{+51}$ kpc}
    \label{side}
\end{figure*}

To find the jump parameter for each of the three cold fronts, we extract a surface brightness profile along the path from the cluster core and through the cold front. We assume spherical geometry and use sectors to fully cover the angular size of the three cold fronts. The sectors used are shown in the right hand panels of Figs \ref{200-220}, \ref{mkreg}, \ref{250_299} and \ref{side} for the inner cold front, the middle cold front, the outer cold front and the {bite region, respectively}. We then fit the surface brightness profiles to a broken powerlaw model, which consists of a powerlaw model jumping to another powerlaw with a different index at a particular jump radius. This model is integrated along the line of sight {using 0.5 arcsec radial bins} and convolved with both Chandra's PSF and a Gaussian to model the width of the cold front. {We calculated the PSF at the location of each cold front using \textit{SAOTRACE 2.1.0} and \textit{MARX 6.0.1}. The model used was also rebinned to 0.5 arcsec to match the bin size of our data.} This method of edge fitting follows the methods used in \cite{Sanders2016b}
and allows us to fit for the width of the cold front, taking into account Chandra's PSF.  
We can also apply these methods across the bite region to get a jump parameter and width measurement. We find our values for the jumps to be 1.67, 1.75, 2.39 and 1.44 for the inner, middle, and outer cold fronts as well as the bite region. We find the widths of these same regions to be 0.73, 6.77, 4.26, and 17.06 kpc. 

To compare these widths with the Coulomb mean free path of particles diffusing from inside the cold front to the outside, given by the expression \citep{Markevitch2007} 

\begin{equation}
      \scriptstyle \lambda_{in\rightarrow out} = \scriptstyle 15 \text{kpc} \left ( \frac{T_{out}}{7 keV} \right )^2\left ( \frac{n_{e,out}}{10^{-3}cm^{-3}} \right )^{-1}\left ( \frac{T_{in}}{T_{out}} \right )\frac{G(1)}{G(\sqrt{T_{in}/T_{out}})}
\end{equation}
we need to measure the temperature and density on either side of the cold fronts.  To find the temperatures of the cold fronts, we extract the spectra along the path of the surface brightness profile using a series of sectors for each cold front (shown in Figs \ref{200-220}, \ref{mkreg}, \ref{250_299} and \ref{side} for the inner cold front, the middle cold front, the outer cold front and the bite region respectively). We then fit the spectra using the methods described in section \ref{spectral}. 

We first obtain projected profiles for the temperature and metallicity, shown by the red points in Figs \ref{200-220}, \ref{mkreg}. We then use {\textit{PROJCT}} in \textsc{xspec} to obtain deprojected profiles for the temperature and metallicity, which are represented by the blue points in Figs \ref{200-220}, \ref{mkreg}. As we expect the outer regions of a cold front have greater temperatures, using deprojection will lower the temperature of the inner part of the cold front. To more accurately determine the deprojected temperatures, we combine bins together to increase the number of photons in each bin before fitting for temperature and abundance. For the outermost cold front and the `bite', we find that the effect of deprojection on the temperature profile is negligible as the outermost gas is very diffuse, so only the projected temperature profile is shown for these in Figs \ref{250_299} and \ref{side}.

{To find the density value, we use the deprojected normalization values using the equation from \cite{Arnaud1996}}:

\begin{equation}
    Norm = \frac{10^{-14}}{4\pi[d_{A}(1+z)]^{2}}\int{n_{e}n_{H}dV}
\end{equation}

{$n_{e}$ is the electron density and $n_{H}$ is the ion density in which $n_{e} = 1.17 n_{H}$ in fully ionized plasma \citep{Grevesse1998}. We assume that the density is constant in each shell and has spherical symmetry. With all of these values, we can solve for the electron density in the ICM at a given radius.}

Table \ref{diffusion1} lists the jump parameters and widths found for each cold front as well as a measurement for the Coulomb mean free path. We see that the observed widths are either less than or consistent with the Coulomb mean free path in all cases, consistent with diffusion being suppressed across all of these edges.

\begin{table}
\begin{tabular}{llllllll}
\hline
\multicolumn{1}{|l|}{\textbf{CF}} & \multicolumn{1}{l|}{\textbf{Jump}} & \multicolumn{1}{l|}{\textbf{kT Jump}} &  \multicolumn{1}{l|}{\textbf{Width (kpc)}} & \multicolumn{1}{l|}{\textbf{MFP (kpc)}} \\ \hline
Inner                             & $1.67^{+0.04}_{-0.03}$                     & $1.81_{-0.60}^{+0.60}$
& $0.73^{+1.31}_{-0.46}$                                     & 0.99                                  \\
Middle                            & $1.75^{+0.07}_{-0.06}$                              & $2.40_{-0.93}^{+0.93}$                                     & $6.77^{+2.58}_{-3.38}$                                     & 6.77                                   \\
Outer                             & $2.39^{+0.09}_{-0.07}$                              & $2.13_{-1.33}^{+1.30}$                                     & $4.26^{+20.90}_{-3.77}$                                     & 85.97                                  \\
Bite                             & $1.44^{+0.16}_{-0.15}$                     & $2.4_{-1.0}^{+1.3}$
& $17.06^{+50.8}_{-15.4}$                                     & 22.38       \\   \hline       
\end{tabular}
\caption{Table of the jump parameters, width of the cold front, and mean free path of the cold front for the three cold fronts in RXJ2014. Notably, we find that the widths of each cold front are systematically lower than or consistent with the mean free path.}
\label{diffusion1}
\end{table}


\begin{figure*}
    \centering
    \hbox{
    \includegraphics[scale=0.45]{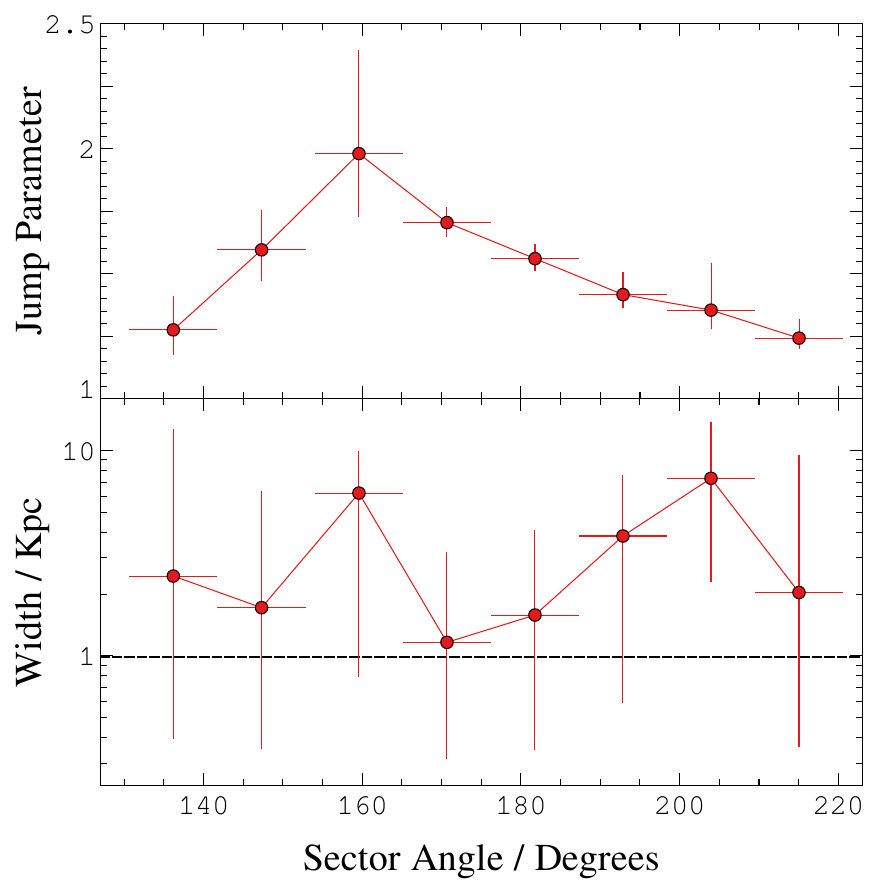}
    \includegraphics[scale=0.35]{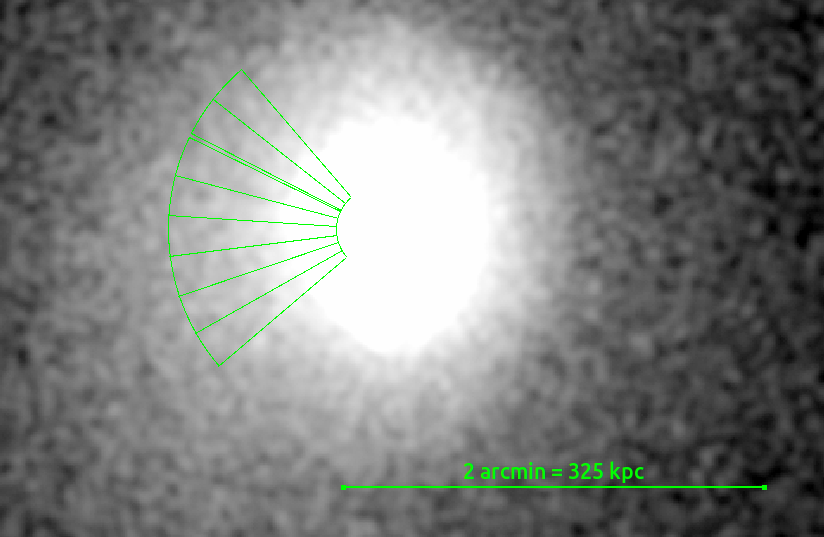}}
    \caption{Left: Profile of jump parameters and widths across the length of the inner cold front using small sectors around 11 degrees in size each. The dashed line denotes the mean free path of the cold front. We show this for the 8 sectors shown in the right hand panel (red points).}
    \label{small8}
\end{figure*}

\begin{figure*}
    \centering
    \hbox{
    \includegraphics[scale=0.5]{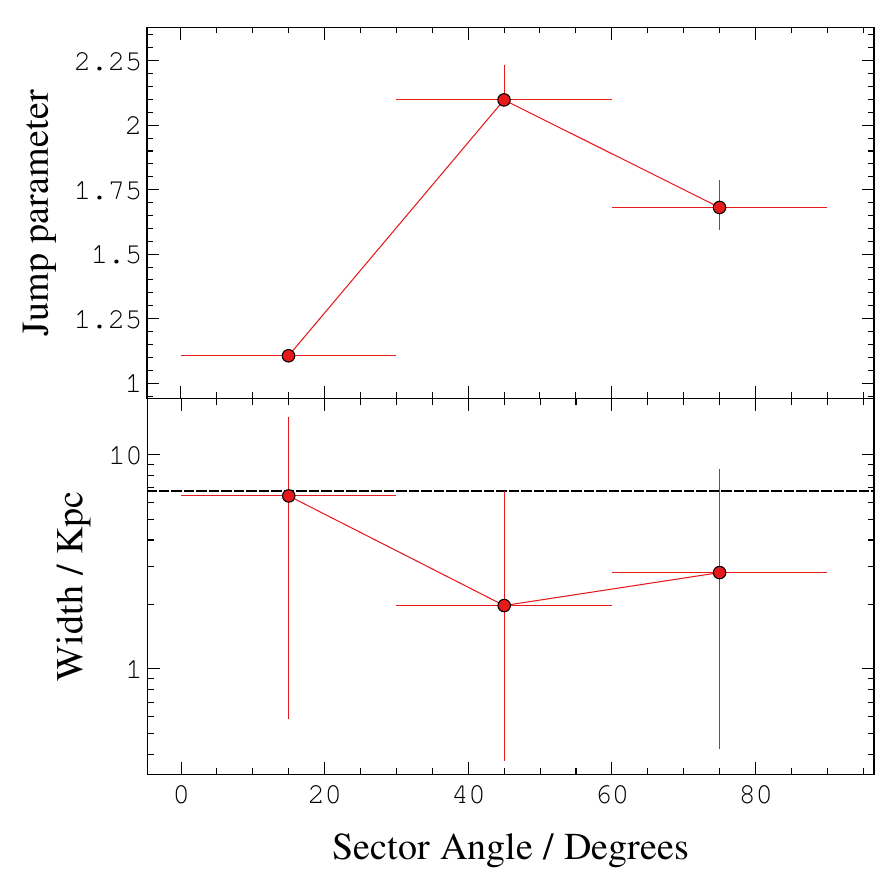}
    \hspace{1.0cm}
    \includegraphics[scale=0.35]{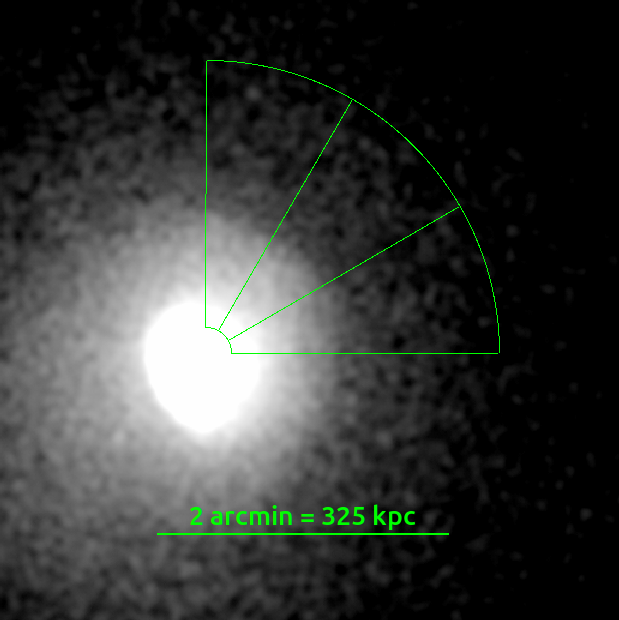}}
    \caption{Left: Profile of jump parameters and widths across the length of the middle cold front using small sectors. The dashed line denotes the mean free path of the cold front. Right: Sectors used in this analysis.}
    \label{middle3}
\end{figure*}

For the inner and middle cold fronts, we also break down the total sectors into smaller sectors to analyze the widths of the cold fronts, similar to \cite{Sanders2016pub} and \cite{Werner2016}. We use 8 sectors of about 11 degrees each across the cold front for the inner cold front, as depicted in Fig \ref{small8}. The azimuthal variation of the jump and width of the inner cold front is shown in Fig \ref{small8} as the red points. For the middle cold front, we use 3 sectors of 30 degrees each from 0 to 90 degrees and plot the results shown in Fig \ref{middle3}. 

The profile of the jump parameter in the inner cold front shows that the jump is highest at the center of the cold front.
For the width profile of the inner cold front, we find that the width is systematically above, but consistent with, the Coulomb mean free path.

 For the middle cold front, the jump profile also shows a sharp peak in the center of the cold front and more smaller jumps towards the edge. We find the width values to be systematically lower than (but consistent with) the mean free path for the middle cold front.

\section{Bite Region}

As mentioned, in sloshing cold fronts there is a potential for KHI to form along the cold front due to the velocity shear between the two regions. These sometimes result in concave regions in the ICM along the swirl, similar to the bite structure in RXJ2014 \citep{Zuhone2016review}. However, a cavity in the ICM inflated by AGN activity is also a plausible scenario for the structure as well. It is typical for strong cool core clusters to have AGN feedback in the center of the cluster, which creates gas cavities in the ICM. However, despite being the strongest cool core cluster in the REXCESS catalog, there is no evidence of any gas cavities either near the core, or having risen in the ICM to the outer regions of the cluster \citep{Mroczkowski2022}. As such, the large concave region in the outer parts of the ICM could potentially be the inner rim of a gas cavity that has risen far away from the core. 

\begin{figure*}
    \centering
    \includegraphics[scale=0.42]{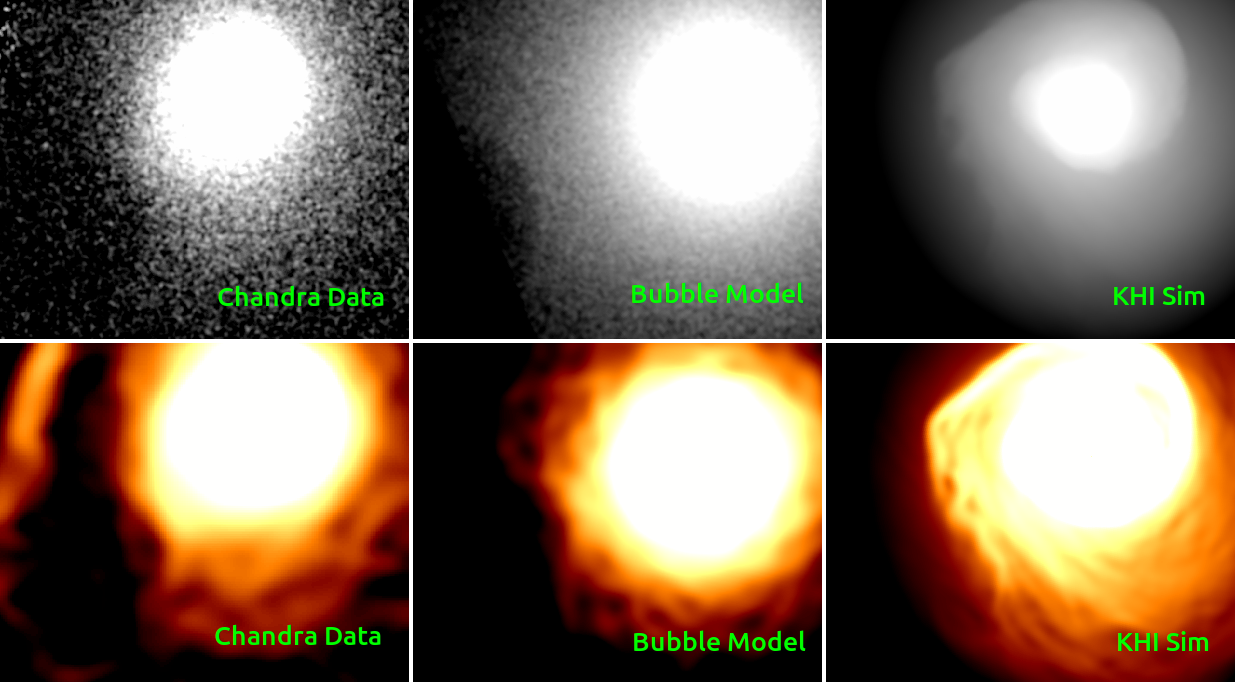}
    \caption{Side by side comparison of the observed concave `bite' feature with simulations of a bubble and a KHI. The top panels show the simulated images compared to the real image, while the bottom panels show the same images put through a GGM filter. The leftmost image is from the Chandra data of RXJ2014. The center image is a bubble model where the bubble is 810 kpc away from the center and has a radius of 332 kpc.  The rightmost image is a frame from the Galaxy Cluster Merger Catalog. This particular frame is of $\beta = 200$ and 4.8 Gyr in which a KHI forms in the ICM. We note that both the bubble model and the KHI model can produce these concave regions similar to our data. }
    \label{simx_comp}
\end{figure*}

Fig \ref{simx_comp} shows similar concave bite shaped structures created from these two scenarios (an AGN inflated bubble and a KHI) compared to our data. To perform this comparison, we model a spherical gas cavity in a uniform cluster, as well as use simulated data of KHI from the Galaxy Cluster Merger Catalog \citep{Zuhone2018b}.


To determine whether the cavity or KHI scenario is preferred, we can compare the surface brightness profiles of these simulations to the observed surface brightness profile across the `bite'. A similar exercise was performed in \cite{Walker2017} for the concave bay in the Perseus cluster, and found that the surface brightness drop over the bay in Perseus was consistent with a KHI from the sloshing simulations of \cite{Zuhone2018b}, and was inconsistent with the drop expected for an AGN inflated bubble of the same size.

\begin{figure*}
    \centering
    \hbox{
    \includegraphics[scale=0.39]{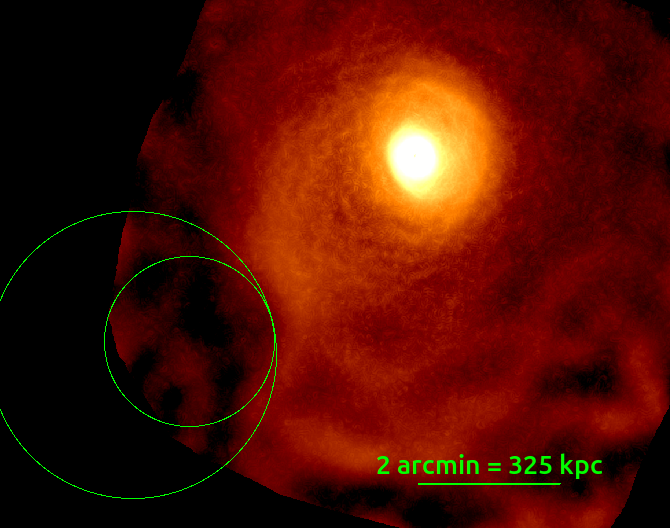}
    \includegraphics[scale=0.49]{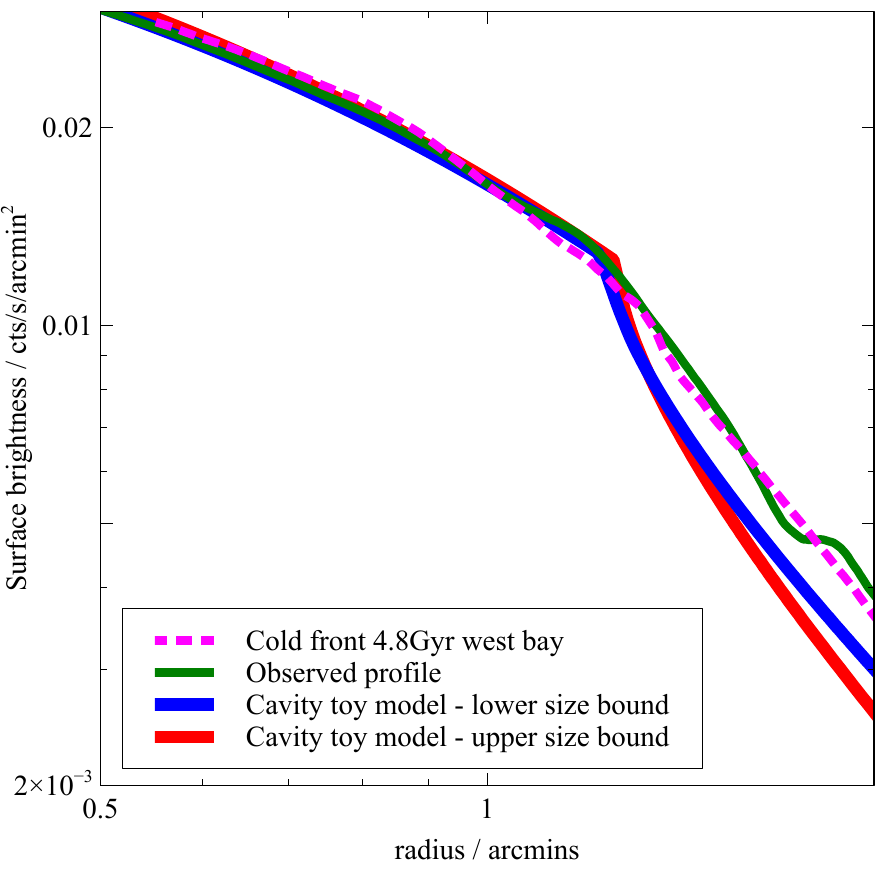}}
    \caption{\textbf{Left:} composite GGM image of RXJ2014. The green circles show the maximum and minimum realistic sizes for a spherical cavity whose inner rim matches the curvature of the `bite'. \textbf{Right:} surface brightness profiles of simulated cavity data, simulated KHI data from the galaxy cluster merger catalog, and the `bite' region. The blue line represents the minimum size cavity with a bubble radius of 197 kpc located 671 kpc away from the core. The red line represents the maximum size cavity with dimensions 810 kpc from the core and a 332 kpc radius. The pink line represents the KHI simulation data, and the green line represents the profile across the actual `bite'. We find that the cavities over predict the jump in the surface brightness, while the KHI data seems to match the slope of the bite region after the jump, which is suggests that the bite is a large scale KHI.}
    \label{bubb}
\end{figure*}

 To do this, we determine the dimensions of a gas cavity that would reasonably fit the bite structure, noting both the radius of the cavity and the distance from the center of the cavity to the cluster core. Since we only see the inner rim (and not its outer rim), we cannot exactly measure the size that the cavity would have. We find that the minimum size spherical cavity that fits the `bite' region would have a radius of 197 kpc, and would have its center 671 kpc from the cluster core. This is shown as the smaller green circle in Fig \ref{bubb}, left. The maximum feasible size for the cavity is shown by the larger circle in Fig \ref{bubb}, left, and has a radius of 332 kpc, with its center being 810 kpc from the cluster core.  We can also calculate the energy output ($PV$) needed to create the minimum and maximum sized cavities. We determine the minimum energy value to be $5.7\times10^{60}$ erg and the maximum value to be $2.7\times10^{61}$ erg.
 
 We then create a model of a uniform cluster with a cavity, using the minimum and maximum possible bubble sizes, and extract surface brightness profiles over them (these simulations are shown in the middle panels of Fig \ref{simx_comp}). Fig \ref{bubb}, right, shows the surface brightness profiles across the two simulated cavities (the blue and red lines), and compares them to the `bite' surface brightness profile (the green line). We find that the cavity models over predict the {drop} of our data, as the jump over the `bite' is not as steep.

We then use the Galaxy Cluster Merger Catalog to simulate the surface brightness profile across a KHI. We use time slice 4.8 Gyr of the $\beta = 200$ simulation from `Sloshing of the Magnetized Cool Gas in the Cores of Galaxy Clusters’ and find a KHI of similar geometry to our system in the outer regions of the simulation (shown in the right hand panels of Fig \ref{simx_comp}), and extract a surface brightness profile from that (shown as the dashed pink line in Fig \ref{bubb}, right). The KHI profile appears to be a good match for the observed `bite' surface brightness profile.

 In \cite{Walker2017}, we find a similar scenario where the toy cavity model over-predicted the jump of the data. The KHI simulation more accurately predicts the slopes of the surface brightness after the jump. This suggests that the concave region in RXJ2014 is due to a large KHI rather than a cavity.

\begin{figure}
    
    \includegraphics[scale=0.5]{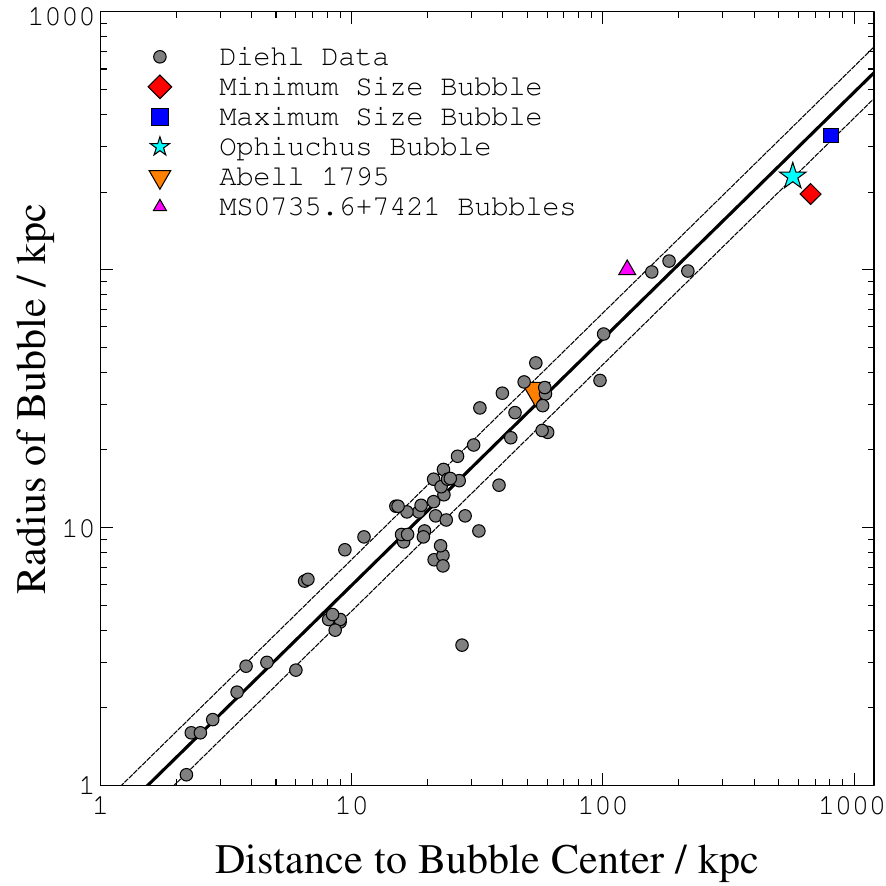}
    
    \caption{Relationship between the radius of the bubbles versus their distance away from the core of the cluster from \citet{Diehl2008}. We compare our minimum and maximum cavity models to cavities in other clusters provided by \citet{Diehl2008}. We fit a best fit line to the Diehl data before adding our data points. Our cavity sizes are broadly consistent with the best fit relation, though we note that the scatter is large. We also add bubbles from the Ophiuchus \citep{Giacintucci2020}, MS0735.6+7421 \citep{McNamara2005nat} and Abell 1795 \citep{Walker2014a1795}.  }
    
    \label{compare}
    
\end{figure}

The relationship between the size of a bubble and the distance it sits from the core has been studied previously \citep{Diehl2008}. It has been found that the variables are related by $R_{b} \propto r^{17/24}$ assuming {the} adiabatic index is $\Gamma = 4/3$, where $R_{b}$ is the distance to the center of the bubble, and $r$ is the radius of the bubble. Using cavity data and best fit relationship provided in \cite{Diehl2008} for other galaxy clusters, we plot both our minimum and maximum bubble sizes to see how they compare, shown in Fig \ref{compare}. Interestingly, both the upper and lower sizes of our cavity are in reasonable agreement with the relationship between the distance to the bubble and the radius of the bubble found for other clusters. However we note that the scatter around this relationship is large.

 To explore this further, we can also compare to other known large cavities. \cite{Giacintucci2020} is an example of a powerful outburst in the Ophiuchus Cluster, noting a cavity roughly 570 kpc away from the core with a bubble radius of around 230 kpc, with a $PV$ of $5\times10^{61}$ erg needed to create this bubble. \cite{McNamara2005nat} is another example of large outbursts in MS0735.6+7421, with a pair of cavities much closer to the center at around 125 kpc, with a bubble radius of 100 kpc and a $PV$ value of $6\times10^{61}$ erg needed to create these bubbles. Our maximum bubble energy of $2.7\times10^{61}$ erg in RXJ2014 is therefore similar in power to the most powerful bubbles ever observed. If the bite region in RXJ2014 is a gas cavity, then even at its lower bound of size it would be one of the furthest cavities from the core of a cluster ever observed. 

 At present, the existing radio data for RXJ2014 does not show any radio emission near the bite region. Future deep radio observations would be needed to determine whether or not the bite region contains radio emitting plasma, which would support the hypothesis that it is an AGN inflated bubble. 

\section{Conclusion}
By analyzing our new deep 144ks observations of the large scale sloshing cluster RXJ2014:

\begin{itemize}
    \item We confirm the location of the three previously observed cold fronts in RXJ2014 using beta model subtraction, GGM filtering, and contour binning. We create a map of the temperatures and metallicities in the cluster.
    \item We find the jump parameters to be 1.67, 1.75, and 2.39 for the cold fronts moving radially outward respectively.
    \item We find the temperature profiles match what we expect for all three cold fronts, with an increase in the value after crossing the interface of the cold front.
    \item We also find the metallicity profile of the inner and middle cold fronts to also conform to expectations, having a higher metallicity value in the cold front, then dropping outside of it.
    \item We find the widths of the cold fronts to be consistent with the Coulomb mean free path within error for each of the cold fronts, with the values for the widths being 0.73 kpc, 6.77 kpc, and 4.26 kpc for the inner, middle and outer cold fronts respectively, indicating that diffusion across the cold fronts is suppressed. 
    \item Small sector analysis shows that the jump parameter is highest in the middle of the inner cold front, and the width values are again consistent with the Coulomb mean free path within error.
    \item We find a large scale concave region in the outskirts of the cluster to the south-east, about 2.6 arcmin (430 kpc) away from the core with a size of about 3.5 arcmin (576 kpc).
    \item The surface brightness profiles of simulated spherical cavities, similar to the size needed to create the concave structure in RXJ2014, over predicts the jump in the observed data. However, the surface brightness profile of the KHI simulation more accurately predicts the slope of the bite region after the jump. 
    \item We match a minimum and maximum size for a cavity to the bite region in the southeastern quadrant of the cluster, with the minimum size being 671 kpc away from the core with a 197 kpc radius, and the maximum size being 810 kpc away from the core with a radius of 332 kpc. We find that the bubbles would be in reasonable agreement with the observed relationship found in other clusters between bubble sizes and distance from the core from \cite{Diehl2008}. The $PV$ value for each of the cavity sizes was calculated to be $5\times10^{60}$ {erg}  for the minimum size cavity and $2.7\times10^{61}$ {erg} for the maximum size cavity. The latter is comparable to the $PV$ values of some of the most powerful bubbles observed such as the ones observed in Ophiuchus and MS0735.6+7421 \citep{Giacintucci2020, McNamara2005nat}.
    \item To better determine the nature of the bite region, deeper radio data of the outskirts of RXJ2014 could determine whether there is a large radio source coming from the bite region, which would be expected if it were an AGN inflated bubble.
\end{itemize}


\section*{Acknowledgements}
SAW acknowledges support from \textit{Chandra} grant GO2-23111X. This work is based on observations obtained
with the \textit{Chandra} observatory, a NASA mission.

\section*{Data Availability}
The \textit{Chandra} Data Archive stores the data used in this paper. The \textit{Chandra} data were processed using the \textit{Chandra} Interactive Analysis of Observations (\texttt{CIAO}) software. The software packages \texttt{HEASoft} and \texttt{XSPEC} were used, and these can be downloaded from the High Energy Astrophysics Science Archive Research Centre (HEASARC) software web page.


\bibliographystyle{mnras}
\bibliography{A3558} 








\bsp	
\label{lastpage}
\end{document}